\documentclass[10pt]{iopart}
\usepackage{graphicx}
\usepackage{amssymb}
\usepackage{iopams}
\usepackage{url}
\usepackage{lineno}

\newcommand{\mudecay}		{\mbox{$\mu^- \rightarrow e^- \nu_{\mu}  \bar{\nu}_{e}$ }}
\newcommand{\weakcap}		{\mbox{$\mu^- \mbox{p} \rightarrow \nu_{\mu}  \mbox{n}$ }}
\newcommand{\egamma}		{\mbox{$\mu^+ \rightarrow e^+ \gamma $ }}
\newcommand{\mueee}		{\mbox{$\mu^\pm \rightarrow e^\pm e^- e^+ $ }}
\newcommand{\muconv}		{\mbox{$\mu^- \mbox{N}\rightarrow e^- \mbox{N} $ }}
\newcommand{\picap}		{\mbox{$\pi^- \mbox{N}\rightarrow \gamma \mbox{N}^* $ }}
\begin{document}
\title{The Mu2e experiment at Fermilab: a search for lepton flavor violation}

\author{
G.~Pezzullo\footnote{on behalf of the Mu2e Collaboration~\cite{MU2ECOL}}}

%% \author{G.~Pezzullo$^{1}$, P.~Murat$^{2}$} 

%% \address{$^1$INFN of Pisa, Pisa, 56123 Italy}
%% \address{$^2$Fermi National Accelerator Laboratory, Batavia, IL 60510 USA}%

\address{INFN sezione di Pisa, Pisa, Italy}
\ead{pezzullo@pi.infn.it}

\begin{abstract}
The Mu2e experiment at Fermilab will search for the charged lepton
flavor violating process of neutrino-less $\mu \to e$ coherent
conversion in the field of an aluminum nucleus.  About $7 \cdot
10^{17}$ muons, provided by a dedicated muon beam line in construction
at Fermilab, will be stopped in 3 years in the aluminum target. The
corresponding single event sensitivity will be $2.5\cdot 10^{-17}$. In
this paper a brief overview of the physics explored by the $\mu \to e$
conversion is given, followed by a description of the Mu2e
experimental apparatus and the expected detector performance.
\end{abstract}

% \pacs{29.40.Mc, 29.40.Vj, 29.30.Dn}
\vspace{2pc}
\noindent{\it Keywords}:  Charged Lepton Flavor Violation Mu2e
%\maketitle
\ioptwocol
\section{Introduction}
In the Standard Model (SM) version where only one Higgs doublet is
included and massless neutrinos are assumed, lepton flavor
conservation is an automatic consequence of gauge invariance and the
renormalizability of the SM Lagrangian. However measurements of the
neutrino mixing parameters during the last decades~\cite{pdg2014}
showed that lepton flavor is not conserved. Including finite neutrino
mass terms in the SM Lagrangian charged lepton flavor violation (CLFV)
is also predicted. CLFV transitions are suppressed by sums over
$\left(\Delta m_{ij}^2/M_W^2\right)^2$, where $\Delta m_{ij}^2$ is
mass-squared difference between the neutrino mass eigenstates $i,\ j$
and $M_W$ is the $W$ boson
mass~\cite{annurev.nucl.58.110707.171126}. Because the neutrino mass
difference is very small ($\Delta m_{ij}^2\leq
10^{-3}$eV$^2$~\cite{pdg2014}) with respect to the W boson mass, the
expected branching ratios reach values below
$10^{-50}$~\cite{annurev.nucl.58.110707.171126,CEI}, which are
unmeasurable by the present facilities. As a consequence, an
observation of CLFV process would represent a clear evidence of new
physics beyond the SM. In the last decades LFV processes were studied
within the supersymmetric extensions of the SM
(SUSY)~\cite{RevModPhys.73.151,annurev.nucl.58.110707.171126} and in
particular within the supersymmetric grand unified theories (SUSY
GUT)~\cite{Barbieri:1994pv, Barbieri:1995tw}. In SUSY models there is
a new source of flavor mixing in the mass matrices of SUSY partners
for leptons and quarks, called sleptons and squarks
respectively. Flavor mixing in the slepton mass matrix would induce
LFV processes for charged leptons. In the SUSY GUT scenario, the
flavor mixing in the slepton sector is naturally induced at the GUT
scale because leptons and quarks belong to the same GUT
multiplet~\cite{RevModPhys.73.151}.  In general, CLFV can be studied
via a large variety of processes: muon decays, such as \egamma,
\mueee, and muon conversion; tau decays: $\tau^\pm \rightarrow \mu^\pm
\gamma$, $\tau^\pm \rightarrow \mu^\pm \mu^+ \mu^-$, etc; meson
decays: $\pi^0\rightarrow \mu e$, $\mbox{K}_{\mbox{\small
    L}}^0\rightarrow \mu e$, $\mbox{K}^+ \rightarrow \pi^+ \mu^+e^-$,
etc; $\mbox{Z}^0$ decays, such as $\mbox{Z}^0 \to \mu e$, etc. The
muon processes have been intensely studied in the CLFV for several
reasons: low energy muon beams can be produced at high-intensity
proton accelerator facilities; Final state of processes in the muon
sector can be precisely measured.  Search for CLFV with muons has been
pursued looking for muon decays (\mbox{$\mu^+ \rightarrow e^+ \gamma
  $} and \mueee), and muon coherent conversion (\muconv). A
model-independent approach represents a convenient way to illustrate
differences among these channels. CLFV can be introduced in the SM by
adding CLF-violating terms to the SM
Lagrangian~\cite{deGouvea:2013zba}:
\begin{eqnarray}\label{eq:lagrangian}
  L_{CLFV} &=& \frac{m_\mu}{(\kappa + 1)\Lambda^2}\bar{\mu}_R\sigma_{\mu\nu}e_L F^{\mu\nu} + \mbox{h.c.} \nonumber\\
  & & + \frac{\kappa}{(\kappa + 1)\Lambda^2}\bar{\mu}_L\gamma_{\mu}e_L \left(\bar{e}\gamma^\mu e\right) + \mbox{h.c.} \ , \nonumber
\end{eqnarray}
where $\Lambda$ is the mass scale of the new physics and $\kappa$ is a
dimensionless parameter. These tow terms in the equation above
correspond to ``dipole'' and ``contact'' interactions terms,
respectively, where $m_\mu$ is the muon mass, $F^{\mu\nu}$ is the
electromagnetic field tensor, and $R$ and $L$ represent the chirality
of the fermion fields.
\begin{figure}[h!]
  \centering
  \includegraphics[width=0.48\textwidth] {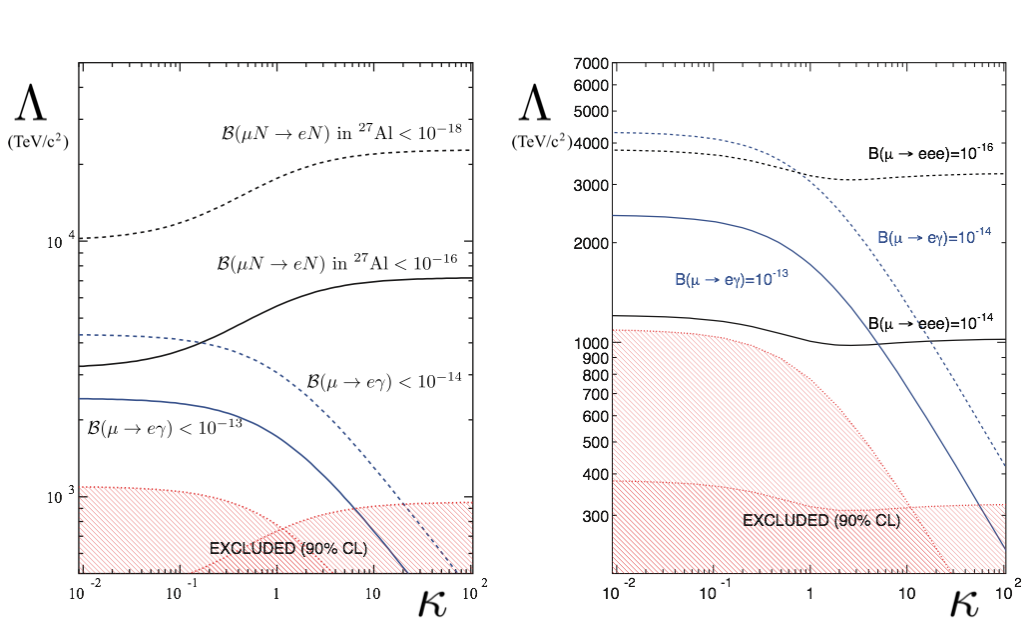}

  \caption{ $\Lambda$ versus $\kappa$ sensitivity plots using: \muconv
    and \egamma (left) and \mueee and \egamma branching ratios. Red
    filled areas represent the region already excluded @ 90 CL.  }
  \label{fig:lambdakappa}
\end{figure}
Figure~\ref{fig:lambdakappa} shows the $\Lambda$ versus $\kappa$
sensitivity plots for the CLFV muon channels. These plots also show
that:
\begin{itemize}
\item
  \muconv search can explore the phase space region where the contact
  term is dominant and \egamma decay is further suppressed;
\item
  The CLFV searches shown here are able to explore new physics mass
  scales significantly beyond the direct reach at the LHC energies.
\end{itemize}
Even if LHC discovers new physics in the second run, precise
measurements of CLFV processes can help discriminate among
several theoretical models~\cite{CEI}.

Experimentally, the search of CLFV using rare muon decays presents
pros and cons. One advantage comes from the fact that these processes
are charge symmetric, so using either $\mu^+$ or $\mu^-$ has no
theoretical disadvantage. However the use of positive muons reduces
significantly the background sources thanks to the absence of capture
processes. Nuclear captures are usually noisy for the detectors
because they produce charged and neutral secondaries: p, n and
$\gamma$. On the other hand, channels involving $\mu^+$ decays
(\egamma, \mueee) suffer for accidental background caused by the
coincidence of two separate processes that can mimic the signal. This
kind of background limits the beam intensity of the experiment.

%%%%%%%%%%%%%%%%%%%%%%%%%%%%%%%%%%%%%%%%%%%%%%%%%%%%%%%%%%%%%%%%%%%%%%%%%%%%%%%%

\section{Experimental searches for \muconv}
When negative muons are stopped in a target (``stopping target'') they
are quickly captured by the atoms ($\sim 10^{-10}$ s) and cascade down
to 1S orbital. Then muons can undergo the following processes: decay
in orbit (DIO) \mudecay; weak capture \weakcap; coherent flavor
changing conversion \muconv. The muon conversion represents a powerful
channel to search for CLFV, because it is characterized by a
distinctive signal consisting in a mono-energetic electron with energy
E$_{\mbox{ce}}$:
$$
  \mbox{E}_{ce}  = \mbox{m}_{\mu} - \mbox{E}_b - \frac{\mbox{E}_\mu^2}{2\mbox{m}_{\rm N}} \ ,
$$
where $\mbox{m}_{\mu}$ is the muon mass at rest, $\mbox{E}_b \sim
\mbox{Z}^2\alpha^2\mbox{m}_{\mu}/2$ is the muonic atom binding energy
for a nucleus with atomic number $\mbox{Z}$, $\mbox{E}_\mu$ is the
nuclear recoil energy, $\mbox{E}_\mu = m_\mu - \mbox{E}_b$, and
$\mbox{m}_{\rm N}$ is the atomic
mass~\cite{annurev.nucl.58.110707.171126}. In case of aluminum, which is the
major candidate for upcoming experiments, E$_{ce}= 104.973$
MeV~\cite{PhysRevD.66.096002}. In muon conversion experiments the quantity:
$$
\mbox{R}_{\mu e}  =  \frac{\Gamma\left( \mu^- + \mbox{N}\rightarrow e^- + \mbox{N} \right)}{\Gamma\left( \mu^- + \mbox{N}\rightarrow \mbox{all captures} \right)}
$$
is measured. The normalization to captures offers a calculation
advantage since many details of the nuclear wavefunction cancel in the
ratio~\cite{PhysRevD.66.096002}.

The coherent conversion leaves the nucleus intact, and there is only
one detectable particle in the final state. The resulting electron
energy stands out from the background (this will be more clear in the
next paragraph), hence muon-electron conversion does not suffer from
accidental background, and extremely high rates can be used.

\subsection{Background sources}
$\mu^-$ stopped in the stopping target can undergo a nuclear
capture~\cite{Measday2001243}. Particles generated in the muon capture
(n, p and $\gamma$) may reach the detector system, and create extra
activity that can either obscure a conversion electron (CE) track or
create spurious hits. As a result, some specific shielding is required
to reduce this background. Additional shielding is required against
cosmic rays that can interact in the apparatus, producing
electrons with an energy mimicking a CE.

\begin{figure}[h!]
  \centering
  \includegraphics[width=0.45\textwidth] {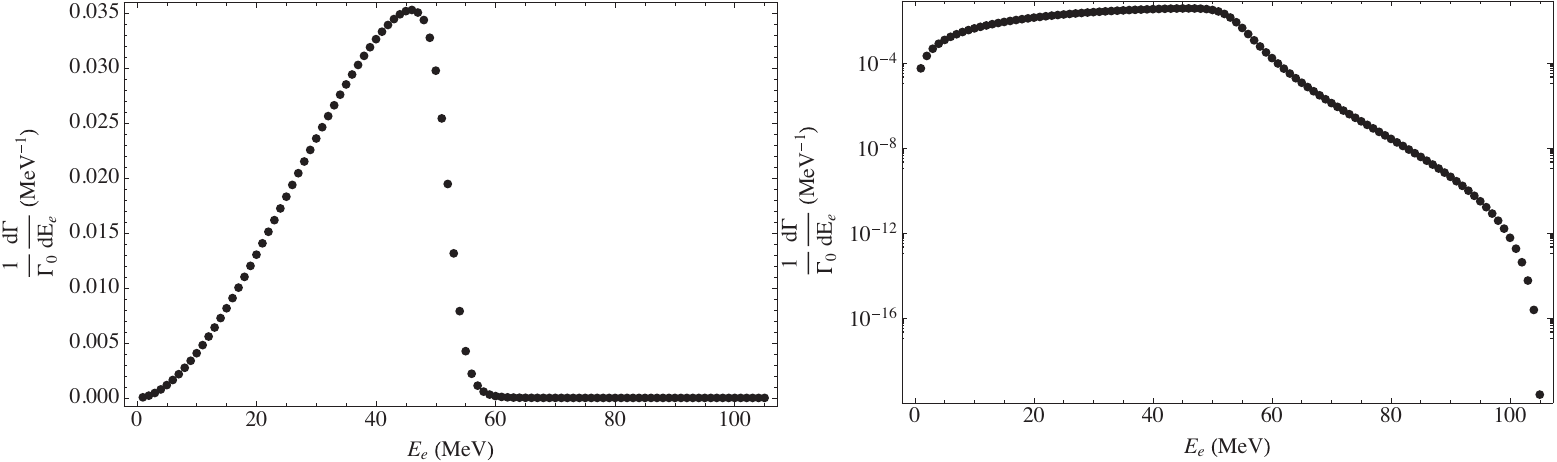}
  \caption{DIO electron energy spectrum on linear (left) and log
    (right) scale, for muons bounded in aluminum nuclei.}
  \label{fig:DIOspectrum}
\end{figure}

Electrons from the high momentum tail of the muon DIO represent the
largest background source for the \muconv
search. Figure~\ref{fig:DIOspectrum} shows the energy spectrum of DIO
electrons~\cite{Czarnecki_Marciano_Tormo_2011}.%%  The endpoint of the spectrum coincides with the conversion
%% electron energy.
%% , and corresponds to the physical case where the
%% neutrinos and the atomic nucleus are perfectly aligned in the opposite
%% direction of the electron.
The main features of the DIO energy spectrum can be summarized as
follows:
\begin{itemize}
\item
  the endpoint of the spectrum corresponds to the energy of the
  electrons from \muconv conversion (CE);
\item
  the overall spectrum is falling as $\left(E_{ce} - E_e\right)^5$, where
  $E_{\rm CE}$ is the CE energy, and $E_e$ is the DIO energy;
\item
  about $10^{-17}$ of the spectrum is within the last MeV from the
  endpoint.
\end{itemize}
Therefore, to reach a sensitivity at the level $O(10^{-17})$ the
detector resolution is crucial.

Another relevant background comes from the radiative pion capture
(RPC) process \picap, followed by the electron-positron pair
conversion of the $\gamma$. Another source of background are pions;
muon beam is generated from low energy protons (below 10 GeV of
energy) interacting with a (production) target, so producing charged
pions that then decay in a transport line. Unfortunately not all pions
decay in the transport line, and, consequently, the muon beam is
contaminated by pion. This source of background is reduced thanks to
the difference between the pion and the bound muon life times. The
pion has a $\tau < $ few tens of ns, while the bound muon has a mean
lifetime of the order of several hundreds of ns (depending on the $Z$
of the material~\cite{Measday2001243}). Therefore using a pulsed beam
structure, it is possible to define a live-gate delayed with respect
to the beam arrival, and to reduce the \picap contribution to the
desired level.
Other beam-related sources of background are: remnant electrons in the
beam that scatter in the stopping target, muon decays in flight, and
antiprotons annihilating in or near the stopping target.

\section{Experimental technique}
\begin{figure}[h!]
  \centering
  \includegraphics[width=0.49\textwidth]{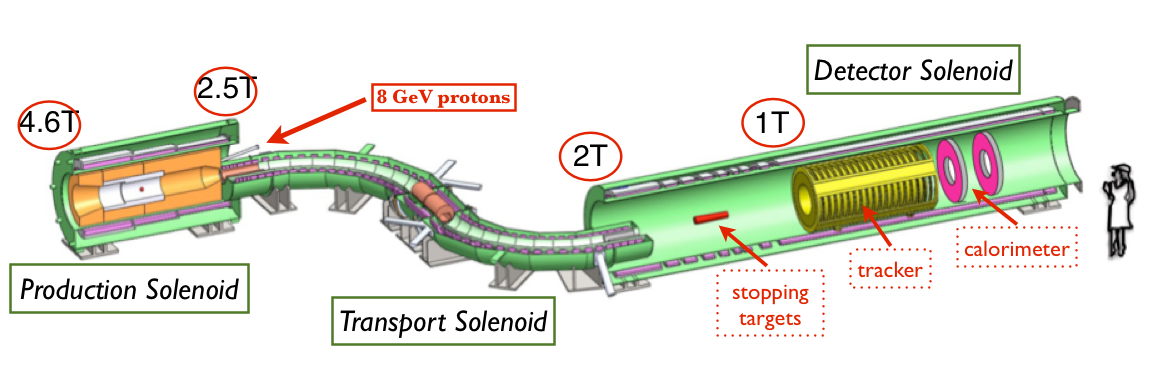}
  \caption{Mu2e apparatus.  }
  \label{mu2e_layout}
\end{figure}

The Mu2e experiment had its genesis back in the 80s, behind the Iron
Curtain. In a way, Mu2e was born in the Soviet Union. In 1989, the
Soviet Journal of Nuclear Physics published a letter to the editor
from physicists Vladimir Lobashev and Rashid Djilkibaev, where they
proposed an experiment that would perform the most thorough search yet
for muon-to-electron flavor violation. In 1992, they proposed the MELC
experiment at the Moscow Meson Factory~\cite{Dzhilkibaev:1995xp}, but
then, due to the political and economic crisis, in 1995 the experiment
shut down. The same overall scheme was subsequently adopted in the
Brookhaven National Laboratory MECO proposal in
1997~\cite{Popp:2001hu}. The Mu2e experimental apparatus includes
three main superconducting solenoid systems:
\begin{itemize}
\item
  {\bf Production solenoid (PS)}, where an 8 GeV pulsed proton beam strikes
  a tungsten target, producing mostly pions;
\item
  {\bf Transport solenoid (TS)}, allowing to select low momenta negative pions
  coming from the production solenoid and letting them to decay into muons
  before they reach the detector region;
\item
  {\bf Detector solenoid (DS)}, housing the aluminum muon stopping target
  and the detector system.
\end{itemize}
Downstream to the proton beam pipe, outside the PS, an extinction
monitor is used to measure the number of protons in between two
subsequent proton pulses. The DS is surrounded by a cosmic ray veto
system, which covers the DS from three sides (the ground is not
covered) and extends up to the midpoint of the TS. Outside the DS, a
stopping target monitor is used to measure the total number of muon
captures. Figure~\ref{mu2e_layout} shows the Mu2e experimental
apparatus.

The Mu2e detector consists of a low-mass straw tube tracker and a
crystal electromagnetic calorimeter. The tracker is assembled of 5 mm
diameter straw tubes. The straws are made of 15 $\mu$m thick metalized
Mylar and have 25 $\mu$m sense wires. The tracker has about 20k straws
combined in 18 tracking stations over a total length of about 3
m~\cite{MU2ETDR}. The straw tubes are orthogonal to the DS axis, and
occupy an annulus with radii from 36 to 70 cm. Two layers of straws
form a panel, 6 panels rotated with respect to each other form a
plane, and a tracking station is made of two rotated planes. Only a
small fraction of the DIO electrons fall into the tracker
acceptance. The inner radius of the tracker planes is such that only
electrons with energies greater than about 53 MeV fall into the
tracker volume: lower energy electrons curl in the solenoidal field
and pass unobstructed through the hole in the center of the
tracker. Because most of the electrons have energy smaller than 60
MeV, a large fraction of them (97\%) do not reach the tracker.  The
momentum resolution is pivotal for eliminating the background, and it
is required to be better than few hundreds of keV/c~\cite{MU2ETDR}.
The calorimeter consists of two disks with an inner (outer) radius of
37.4 (66) cm, and a relative distance of 75 cm. Each disk is composed
of about 600 pure CsI crystals read out by Silicon
Photomultipliers. The crystal size is $3.4\times 3.4\times 20$
cm$^3$. Simulation studies and beam tests with a reduced scale
prototype~\cite{MU2EPHDTHESIS} showed that the calorimeter performance
for 100 MeV electrons are: a timing resolution of about 100 ps, and an
energy resolution of about 5\%.
Calorimeter information allows to improve track reconstruction, and to
provide a particle identification tool. The calorimeter may also be
used to trigger high energy electron candidates, reducing the
throughput of the data acquisition system.
A cosmic ray veto system is also present to veto atmospheric muons
that can interact in the DS, generating fake CE candidates.
\begin{figure}[h!]
  \centering
  \includegraphics[width=0.49\textwidth]{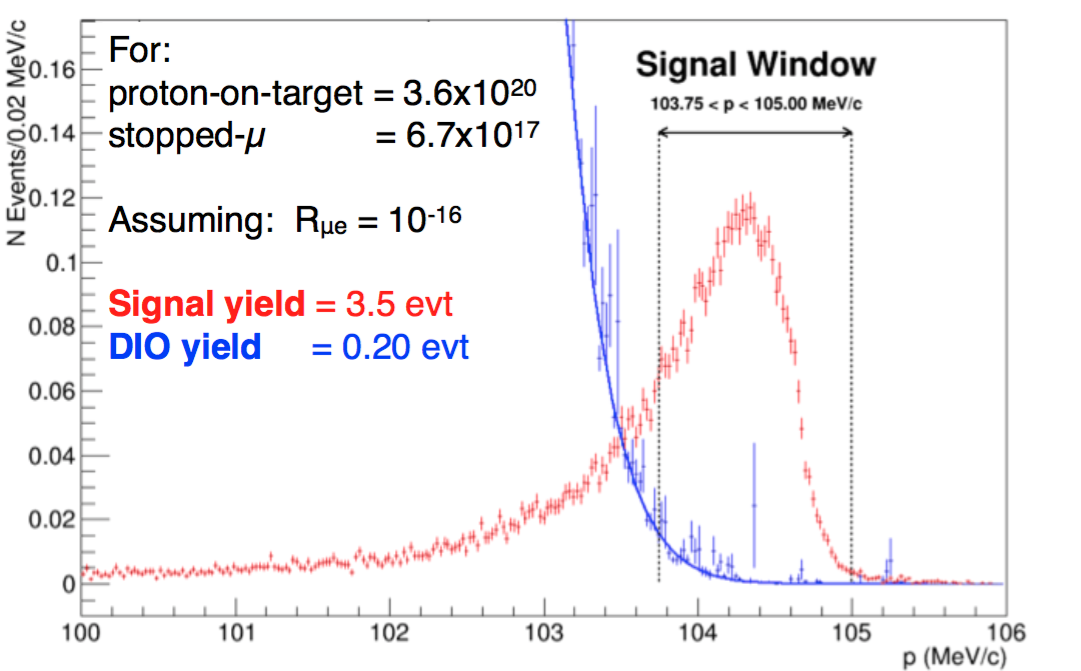}
  \caption{Signal and DIO background yield normalized to 3 years of
    Mu2e data taking. }
  \label{CEYIELD}
\end{figure}
Figure~\ref{CEYIELD} shows the signal and the DIO background yield,
normalized to 3 years of data taking. Same Figure shows that if we
define a narrow momentum window (signal window) around the signal peak,
the mean expected background from the DIO electrons is about 0.2
events, while assuming $\rm R_{\mu e}=10^{-16}$ the signal yield is expected to
be about 3.5 events. It has also been shown in reference~\cite{MU2ETDR}
that the contribution from the other background sources adds 0.3 events
in the signal window.

\section{Summary}
The Mu2e experiment will search for the $\mu \to e$ conversion in the
field of an aluminum nucleus with a single event sensitivity of
2.9$\cdot 10^{-17}$. This will improve the current best limit by 4
orders of magnitude, probing new physics at scales up to 10,000
TeV. The detector system consists of a low-mass straw tube tracker
that will measure the signal momentum with an expected resolution
better than 200 keV/c, and a crystal calorimeter made of pure CsI that
will measure the energy (time) of the signal particle with a
resolution of about 5\% (100 ps). The design of the apparatus is
mature and the construction of several components is underway to start
data taking in the end of 2020.

\section*{Acknowledgments}
We are grateful for the vital contributions of the Fermilab staff and
the technical staff of the par- ticipating institutions. This work was
supported by the US Department of Energy; the Italian Istituto
Nazionale di Fisica Nucleare; the US National Science Foundation; the
Ministry of Education and Science of the Russian Federation; the
Thousand Talents Plan of China; the Helmholtz Association of Germany;
and the EU Horizon 2020 Research and Innovation Program under the
Marie Sklodowska-Curie Grant Agreement No.690385. Fermilab is operated
by Fermi Research Alliance, LLC under Contract No. De-AC02-07CH11359
with the US Department of Energy.

%% References

%% \nocite{*}
%% \bibliographystyle{elsarticle-num}
%% \bibliography{jos}

\section*{References}
\begin{thebibliography}{1}
\bibitem{MU2ECOL} Mu2e Collaboration, http://mu2e.fnal.gov/collaboration.shtml
\bibitem{pdg2014} Olive, K. and others, \emph{Review of Particle
  Physics}, Chin. Phys., C38, 2014
\bibitem{annurev.nucl.58.110707.171126} Marciano, W. J.  and others,
  \emph{Charged Lepton Flavor Violation Experiments}, Annual Review of
  Nuclear and Particle Science, 58, 1, 315-341, 2008
\bibitem{CEI} Cei, F. and Nicolo, D., \emph{Lepton Flavour Violation
  Experiments}, Adv. High Energy Phys., 2014, 2014
\bibitem{RevModPhys.73.151} Kuno, Y. and Okada, Y.,\emph{Muon decay and
  physics beyond the standard model}, Rev. Mod. Phys., 73, 1, 151-202,
  2001
\bibitem{Barbieri:1994pv} Barbieri, R. and Hall, L. J., \emph{Signals
  for supersymmetric unification}, Phys. Lett., B338, 1994
\bibitem{Barbieri:1995tw} Barbieri, R. and Hall, L. J. and Strumia,
  A., \emph{Violations of lepton flavor and CP in supersymmetric
    unified theories}, Nucl. Phys.", B445, 1995
\bibitem{deGouvea:2013zba} de Gouvea, A. and Vogel, P., \emph{Lepton
  Flavor and Number Conservation, and Physics Beyond the Standard
  Model}, Prog. Part. Nucl. Phys., 71, 2013
\bibitem{PhysRevD.66.096002} Kitano, R. and Koike, M. and Okada, Y.,
  \emph{Detailed calculation of lepton flavor violating muon electron
    conversion rate for various nuclei}, Phys. Rev., D66, 2002
\bibitem{Measday2001243} Measday, D. F., \emph{The nuclear physics of
  muon capture}, Phys. Rept., 354, 2001
\bibitem{Czarnecki_Marciano_Tormo_2011} Czarnecki, A. and Garcia i
  Tormo, X. and Marciano, W.J., \emph{Muon decay in orbit: Spectrum of
    high-energy electrons}, Phys. Rev., D84, 1, 8, 2011
\bibitem{Dzhilkibaev:1995xp} Dzhilkibaev, R. M. and Lobashev, V. M.,
  \emph{The solenoid muon capture system for the MELC experiment},
  Beam Dynamics and Technology Issues for mu+ mu- Colliders:
  Proceedings of the 9th ICFA Advanced Beam Dynamics Workshop,
  Oct. 15-20 1995, Montauk, New York, 1995
  
\bibitem{Popp:2001hu} Popp, James L., \emph{The MECO experiment: A
  Search for lepton flavor violation in muonic atoms},
  Nucl. Instrum. Meth., A472, 354-358, 2000
\bibitem{MU2ETDR} Bartoszek, L. and others, \emph{Mu2e Technical
  Design Report}, arXiv:1501.05241, [physics.ins-det], 2014
\bibitem{MU2EPHDTHESIS} Pezzullo, G., \emph{The Mu2e crystal
  calorimeter and improvements in the $\mu^-\mbox{N} \to e^-\mbox{N}$
  search sensitivity}, FERMILAB-THESIS-2016-02, 2016
  
\end{thebibliography}



\end{document}